\documentclass{cjaa}                    

\usepackage{graphicx}                   
\input{epsf.sty}                        
\input{psfig.sty}                       

\begin{document}

   \title{Fireball and Cannonball Models
          of Gamma-Ray Bursts\\ Confront Observations}

   \author{Arnon Dar} 
      
\offprints{Arnon Dar} 

   \institute{Physics Department and 
Space Research Institute\\
    Technion - Israel Institute of Technology, Haifa 32000, Israel
             \email{arnon@physics.technion.ac.il}
     }

   \date{Received~~2005 September 27; accepted~~2005~~month day}

   \abstract{The two leading contenders for the theory of gamma-ray bursts 
(GRBs) and their afterglows, the Fireball and Cannonball models, are 
compared and their predictions are confronted, within space limitations, 
with key observations, including recent observations with SWIFT.}

   \authorrunning{Arnon Dar}
   \titlerunning{The Fireball and Cannonball Models of GRBs-Comparison} 


\maketitle 
\section{Introduction} 

\noindent 
The currently best-studied theories of Gamma-Ray Bursts (GRBs), 
X-Ray Flares (XRFs) and their afterglows (AGs) are the {\it Fireball} (FB) 
models (see, e.g. Zhang \& M\'esz\'aros~\cite{ZM2004} and 
Piran~\cite{Pira2005} for recent reviews) and the {\it Cannonball} (CB) 
model (see, e.g., Dar \& De R\'ujula~ \cite{DD2000},\cite{DD2004}; Dado, 
Dar \& De R\'ujula~\cite{DDD2002},\cite{DDD2003} and references therein).  
In spite of their similarly sounding names, these two models are (or were 
initially) completely different in their basic hypothesis, in their 
description of the data, and in their predictions.  While FB models 
are/were generally welcomed by the GRB community (see however, 
Dermer~\cite{Derm2002}), the CB model was not. In fact, many underlying 
ideas of the CB model, which were proved to be correct and even 
path-breaking for the GRB field, met skepticism, strong opposition, and 
initial dismissal.  These ideas included the collimated nature of 
GRBs/XRFs and their afterglows, the SN/XRF-GRB association and the trivial 
unification of GRBs and XRFs, which were all suggested in print long 
before they were adapted in the FB models. GRBs/XRFs and their AGs 
are notorious for their diversity. Yet, both models claim to predict 
correctly their main properties. No doubt, both models are a 
simplification of a very complex phenomenon and may require modifications 
when compared with future observations.  But their assumptions and 
predictions should be falsifiable and should be tested not against 
prejudices of astrophysicists, but against key observations, as I shall 
try to do here within space limitations.

\section{The FB and CB Models} 

\begin{figure}
   \vspace{2mm}
   \begin{center}
   \hspace{3mm}
\psfig{figure=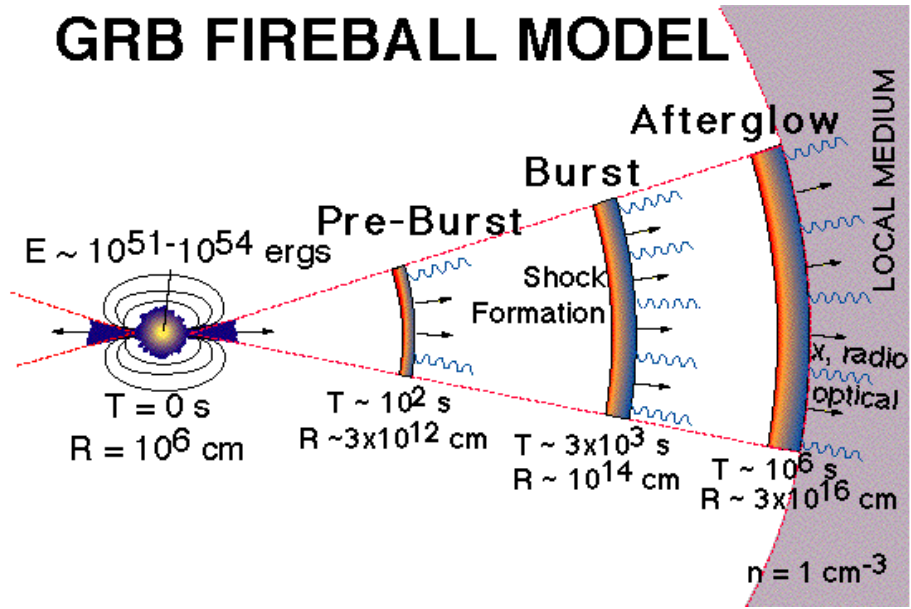,width=120mm,height=80mm,angle=0.0} 
\caption{Illustration of the FB model and 
its internal/external shock scenario, adapted from 
Ghisellini~\cite{Ghis2001}.} 
\label{FBmodel} 
\end{center} 
\end{figure}

\noindent 
The current FB models of GRBs evolved a 
long way from the first spherical FB models, suggested by 
Paczynski ~(\cite{Pacz1986}) and Goodman~(\cite{Good1986})\footnote{
See, e.g., Shemi \& Piran~\cite{SP1990};  Narayan, 
Paczynski \& Piran~\cite{NPP1992};  M\'esz\'aros \& Rees ~\cite{MR1992}; 
M\'esz\'aros \& Rees~\cite{MR1993}; Paczynski \& Rhoads~\cite{PR1993};  
Katz~\cite{Katz1994a},\cite{Katz1994b}; M\'esz\'aros \& 
Rees~\cite{MR1997}; Waxman~\cite{Waxm1997a},\cite{Waxm1997b}; Dermer \& 
Mitman~\cite{DM1999} and the more recent reviews by 
Piran~(\cite{Pira1999},\cite{Pira2000}), M\'esz\'aros~(\cite{Mesz2002}), 
Hurley, Sari \& Djorgovski~(\cite{HSD2002}), Waxman~(\cite{Waxm2003a}), 
Zhang \& M\'esz\'aros~(\cite{ZM2004}) and Piran~(\cite{Pira2005}).}. The 
most popular ones (see e.g. Zhang \& M\'esz\'aros~ \cite{ZM2004} 
and Piran~\cite{Pira2005}) assume that GRBs are emitted from highly 
relativistic conical fireballs (M\'esz\'aros \& Rees~\cite{MR1992}; 
Levinson \& Eichler~\cite{LE1993}, Woosley~\cite{Woos1993a,Woos1993b}) 
produced in hypernovae - an hypothesized rare class of superenergetic 
supernovae (SNe) type Ic, generated by direct collapse of massive stars to 
black holes (Paczynski~\cite{Pacz1988};
Iwamoto et al.~\cite{Iwam1998}; MacFadyen \& 
Woosley~\cite{MW1999})\footnote{Originally, 
Woosley~(\cite{Woos1993a,Woos1993b}) argued against a GRB-SN association 
and suggested that GRBs are produced by ``Failed Supernovae'', i.e., by 
collapsars which do not produce a supernova. After the discovery of the 
GRB980425/SN1998bw association by Galama et al.~(\cite{Gala1998}), the 
``Failed Supernovae'' became ``Hypernovae'', i.e. super-energetic 
supernovae (Iwamoto et al.~\cite{Iwam1998}; MacFadyen \& 
Woosley~\cite{MW1999}).}. The GRB pulses are assumed to be produced by 
synchrotron emission from collisions between highly 
relativistic conical shells ejected in the hypernova explosion, while the 
GRB afterglow is assumed to be synchrotron radiation emitted when the 
merged shells collide with the interstellar medium (ISM) and drive a 
forward blast wave into the ISM and a reverse shock into the merged 
shells. The FB model is illustrated in Fig.~\ref{FBmodel} adapted from 
Ghisellini~\cite{Ghis2001}. 

\begin{figure}
   \vspace{10mm}
   \begin{center}
   \hspace{3mm}
\psfig{figure=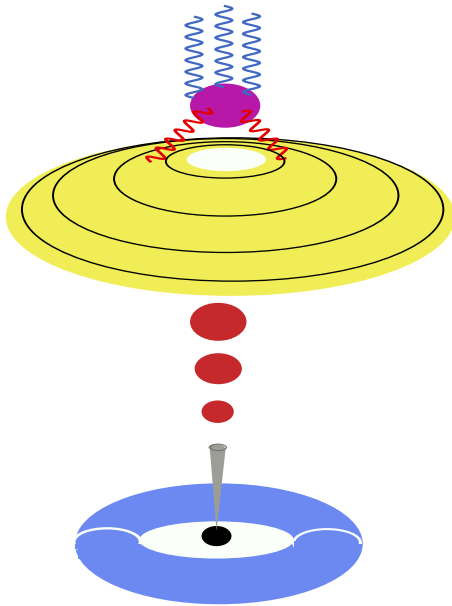,width=60mm,height=80mm,angle=0.0} 
\caption{A fewer-kbyte version of 
Fig.~(3) of Dar \& De R\'ujula (2000a) showing an ``artist's view'' (not 
to scale) of the CB model of GRBs and their afterglows. A core-collapse SN 
results in a compact object and a fast-rotating torus of non-ejected 
fallen-back material. Matter (not shown) abruptly accreting into the 
central object produces a narrowly collimated beam of CBs, of which only 
some of the ``northern'' ones are depicted. As these CBs move through the 
``ambient light'' surrounding the star, they Compton up-scatter its 
photons to GRB energies.} 
\label{figCB} 
\end{center} 
\end{figure}

\noindent 
The CB model (Dar \& De R\'ujula~\cite{DD2000},\cite{DD2004}; Dado, Dar 
\& De R\'ujula~\cite{DDD2002},\cite{DDD2003}) is an elaboration on the 
ideas of De R\'ujula~(\cite{ADR1987}), Shaviv and Dar~(\cite{SD1995}), 
Dar~(\cite{Dar1997}, \cite{Dar1998}) and Dar \& Plaga~(\cite{DP1999}). In 
the CB model, {\it long-duration} GRBs and their AGs are produced by 
bipolar jets of CBs which are ejected in {\it ordinary core-collapse} 
supernova explosions. An accretion disk or torus is hypothesized to be 
produced around the newly formed compact object, either by stellar 
material originally close to the surface of the imploding core and left 
behind by the explosion-generating outgoing shock, or by more distant 
stellar matter falling back after its passage (De R\'ujula~ 
\cite{ADR1987}). As observed in microquasars (e.g.~Mirabel \& Rodriguez 
\cite{MiRo1999}; Rodriguez \& Mirabel~\cite{RoMi1999} and references 
therein), each time part of the accretion disk falls abruptly onto 
the compact object, a pair of CBs made of {\it ordinary atomic matter} are 
emitted with high bulk motion Lorentz factors, $\gamma$, in opposite 
directions, along the rotation axis, where matter has already fallen back 
onto the compact object due to lack of rotational support. The 
$\gamma$-rays of a single pulse in a GRB are produced as a CB coasts 
through the SN glory - the SN light scattered by the SN and pre-SN ejecta. 
The electrons enclosed in the CB Compton up-scatter photons to GRB 
energies. Each pulse of a GRB corresponds to one CB. The timing sequence 
of emission of the successive individual pulses (or CBs) in a GRB reflects 
the chaotic accretion process and its properties are not predictable, but 
those of the single pulses are (Dar \& De R\'ujula~\cite{DD2004}). An 
artist's view of the CB model is given in Fig.~\ref{figCB}.

\noindent 
The rapid expansion of the CBs stops shortly after ejection by 
their interaction with the ISM (Dado, Dar \& De R\'ujula~\cite{DDD2002}). 
During this rapid expansion and cooling phase their afterglow emission is 
usually dominated by thermal bremsstrahlung (TB) and line emission. Later, 
their emission is 
dominated by synchrotron radiation from swept-in ISM electrons spiraling 
in the CBs' enclosed magnetic field.\\

\noindent 
Table~\ref{FBCBGRB1} compares the main assumptions of the FB 
and CB models on the production of GRBs, while Table~\ref{FBCBAG1} 
compares them for AGs. 
\begin{table}[] 
\caption[]{Main Assumptions of the FB and CB Models of Long Duration GRBs}
  \label{FBCBGRB1}
  \begin{center}
  \begin{tabular}{lclcl}
  \hline\noalign{\smallskip}

Property & Fireball Model & Cannonball Model\\

\hline\noalign{\smallskip}

{\bf Progenitors} & Massive Stars & Massive Stars, Compact Binaries\\

{\bf Event} & Hypernova,  & CC Supernova, AIC of wd/ns in a binary \\

{\bf Environment} & Progenitor Wind, ISM & SN+Pre SN ejecta, SB\\

{\bf Remnant} & bh & bh, or qs, hs or ns \\

{\bf Ejecta} & Baryon poor $e^+e^-$ Shells & Ordinary-Matter Plasmoids \\

{\bf Geometry} & Conical Shells & Cannonballs (CBs) \\

{\bf Ejection time} & During Core Collapse & After Fall-Back of Ejecta\\

{\bf Radiation} & Synchrotron from Colliding Shells& ICS of SN Glory by 
CBs\\

{\bf Lorentz Factor} & $\Gamma > 300$ & $\Gamma\sim 1000$\\

{\bf Viewing Angle} & $ \theta<\theta_j$ & $\theta \sim 1/\Gamma $ 
\\

\noalign{\smallskip}
\hline

\end{tabular} 
\end{center} 
{\bf Abbreviations}: bh $=$ black hole; ns 
$=$ neutron star; qs $=$ quark star; hs $=$ hyper star; wd $=$ white 
dwarf;\\
SN $=$ supernova; SB $=$ Superbubble;  AIC $=$ accretion induced collapse; 
ICS $=$ inverse Compton scattering; 
$\theta_j=$ opening angle of the conical jet 
(relative to its axis). 
\end{table}

\begin{table}[] 
\caption[]{The FB and CB Model Assumptions for AGs of Long Duration GRBs} 
  \label{FBCBAG1}
  \begin{center}\begin{tabular}{lclc}
  \hline\noalign{\smallskip}

Property & Fireball Model & Cannonball Model\\

\hline\noalign{\smallskip} 
{\bf Origin} & Blast wave in ISM & CB Interaction with ISM\\

{\bf Ejecta} & Conical $e^+e^-$ shells  & Jet of ordinary matter 
plasmoids\\

{\bf External Medium} & Progenitor wind, ISM & Progenitor wind 
$\rightarrow$ SB $\rightarrow$ ISM\\

{\bf Early AG} & SR from reverse shock& Brem. and line 
cooling\\

{\bf Late AG} & SR from shocked ISM & SR from CBs and scattered ISM\\

{\bf Lightcurve Break} & Jet expansion+deceleration & Off-axis viewing 
of decelerating CBs\\

{\bf Early Flares} & Extended central activity & Late accretion episodes\\

{\bf Late Flares} & Extended central activity & Encounter with density 
bumps\\

{\bf Dark Bursts} & Circumburst absorption & Circumburst absorption \\

\noalign{\smallskip}\hline

\end{tabular} 
\end{center} 
{\bf Abbreviations:} SR $=$ Synchrotron Radiation; Brem. $=$ Bremsstrahlung
\end{table}

\section{FB and CB Models Confronting GRB Data}

Despite their diversity GRBs have 
a series of common features: The typical energy of their $\gamma$ rays is 
a fraction of an MeV.  The energy distributions are well described by a 
``Band spectrum'', with ``peak energies'' spanning a surprisingly narrow 
range. The time structure of a GRB consists of pulses, superimposed or 
not, rising and decreasing fast. The number of photons in a pulse, the 
pulses' widths and their total energy vary within broad but specific 
ranges. Within a pulse, the energy spectrum softens with increasing time. 
The duration of a pulse decreases at higher energies and its peak 
intensity shifts to earlier time. Many other correlations between pairs of 
observables have been identified. Last, based on three measured events, 
the $\gamma$-ray polarization seems to be very large. A satisfactory 
theory of GRBs should naturally predict/explain these properties. 
Table~\ref{FBCBGRB2} summarizes which GRB properties were correctly 
predicted/explained by the FB and CB models. Because of space limitation, 
I shall discuss briefly only a few entries in Table~\ref{FBCBGRB2}. 
\begin{table}[] 
\caption[]{Comparison between Falsifiable Predictions of 
the FB and CB Models and Observational Data on Long GRBs.}
  \label{FBCBGRB2}
  \begin{center}
  \begin{tabular}{lclcl}
  
\hline\noalign{\smallskip}

Property & Fireball Model & Cannonball Model & Observations\\

\hline\noalign{\smallskip}

{\bf ``Peak'' $\gamma$-Ray Energy} & NP & SP: Eq.~(\ref{Epeak1})  &$\sim $ 
250 keV\\

{\bf Typical Duration} ($T_{90}$) & NP & NP & 30s\\

{\bf Mean No of Pulses} ($n_p$) & NP & NP &$\sim 6$\\

{\bf Pulse Shape} & FP &  SP: Eq.\ref{pheno} ``FRED'' &``FRED''\\
 
{\bf ``Isotropic Energy''/Pulse} & NP & SP: $0.8\times 10^{53}\, \delta_3^3$ 
ergs & $10^{48}-10^{54}$\, ergs\\

{\bf FWHM of Pulses } & FP & SP: $\sim 
0.5\,(1+z)/\delta_3$ s & 1/2-200 s& \\

{\bf Pulse Spectrum} & FP & SP: Thermal Brem. + PL Tail & ``Band'' 
spectrum\\

{\bf Pulse Spectral Evolution} & NP & SP: Eq.~(\ref{Tevol})& Hard to soft 
\\

{\bf Scintillations} & NP & NP & Debated \\

{\bf Polarization} & FP & SP: Eq. (8) & $ >50\%$ (3 GRBs)\\

{\bf Correlations (Pulses):} & & & \\

$ (1+z)\, E_p(E^{iso}_\gamma) $ & NP & SP: $\sim 
[E^{iso}_\gamma]^{1/3-1/2}$ & 
$\sim [E^{iso}_\gamma]^{0.46\pm 0.13}$\\

{\bf FWHM($E_\gamma$)} & SP: $\sim E_\gamma^{-1/2}$ & SP $\sim 
E_\gamma^{-1/2}$& 
$\sim E_\gamma^{-1/2}$\\

{\bf Rise-Time/FWHM}& NP &SP: $\sim 0.27$ & $\sim 0.30$\\

\noalign{\smallskip}\hline

\end{tabular} 
\end{center} 
{\bf Abbreviations}: NP=Not Predicted; 
FP=Failed Prediction; SP=Successful Prediction\\
PL=Power Law; FRED=Fast Rise Exponential Decay; 
$\delta_3\times=\delta/10^3$ . 
\end{table}\\

\noindent 
{\bf 3.1 Peak Photon Energy, Isotropic Equivalent Energy and the 
``Amati Relation''}\\ 
No simple explanation/predictions for the peak photon energy, 
the isotropic equivalent energy or the
``Amati Relation'' has been provided by the FB model.\\ 
In the CB model (e.g. Dar \& De 
R\'ujula~\cite{DD2004}), the observed peak-energy of $\gamma$-rays 
produced 
at a redshift $z$ by ICS of SN glory with a thermal bremsstrahlung 
spectrum $dn_\gamma/dE \sim E^{-\alpha}\, exp(-E/T) $ with a typical 
energy $\epsilon_\gamma\sim T\sim 1\, eV$ and a typical $\alpha\sim 1\,,$ 
is given by, 
\begin{equation}
 E_p\approx (2-\alpha)\, \gamma\,\delta\,T/(1+z)\,, 
\label{Epeak1} 
\end{equation} 
where, 
$\gamma\sim 1000 $ is the initial Lorentz factor of a CB, $\theta\ll 1 $ 
is 
the viewing angle relative to the CB motion and $\delta=1/\gamma\, 
(1-\beta\, \cos\theta)\approx 2\, \gamma/(1+\gamma^2\, \theta^2)$ is its 
Doppler factor. For the typical viewing angle, $\theta\sim 1/\gamma\, ,$ 
and the mean redshift, $<z>=2.75\, $ of SWIFT GRBs, $E_p\sim 265$ keV, in 
agreement with the GRB observations of BATSE, BeppoSAX and SWIFT.
 
\noindent 
Under the assumption of isotropic emission in the CB rest frame, 
Doppler boosting and relativistic beaming yield a $\gamma$-ray fluence 
$F_\gamma$ of a GRB pulse, which is proportional to $\gamma\, \delta^3\, ,$ 
\begin{equation} 
F_\gamma\approx \delta^3\, [(1+z)\, E'_\gamma/ 4\, \pi\, D_L^2]\,, 
\label{Fluence} 
\end{equation} 
where $E'_\gamma\sim 
N_\gamma\, \gamma\, T $ is the total energy in the CB rest-frame of the 
$N_\gamma$ ambient photons which suffer Compton scattering in the CB. 
Consequently, under the assumption of isotropic emission in the observer 
frame, the inferred `GRB isotropic $\gamma$-ray energy' in a GRB pulse, is 
\begin{equation}
E^{iso}_\gamma= 4\, \pi\, D_L^2\, F_\gamma/(1+z)
                  \approx \delta^3\, E'_\gamma\, . 
\label{Eiso} 
\end{equation} 
In the CB 
model $E'_\gamma\approx 0.8\times 10^{44}$ ergs (Dar \& De 
R\'ujula~\cite{DD2004}) which yields $E^{iso}_\gamma[pulse]\sim 0.8\times 
10^{53}$ ergs for a single pulse, or $E^{iso}_\gamma[GRB]\sim 5\times 
10^{53}$ ergs for the typical values, $\theta\sim 1/\gamma$ and 
$\gamma\sim 1000\, ,$ and 6 pulses in a GRB, in agreement with 
observations.
 
\begin{figure}
   \vspace{2mm}
   \begin{center}
   \hspace{3mm}
\psfig{figure=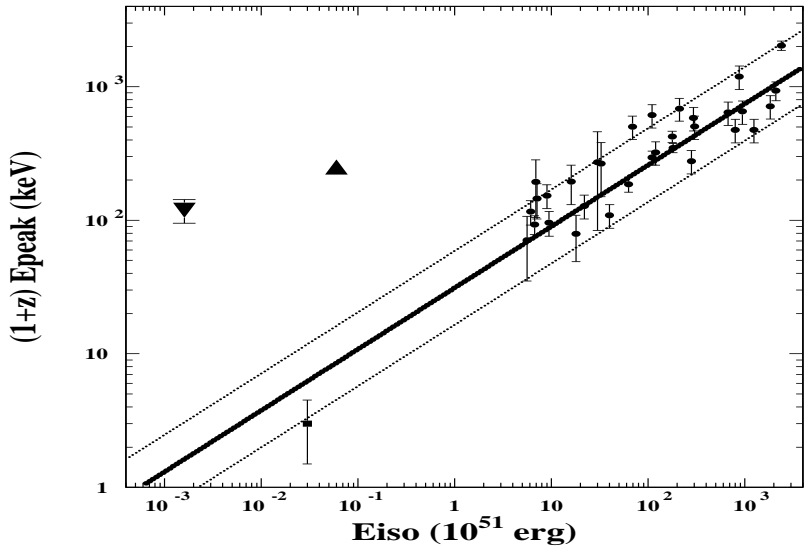,width=120mm,height=80mm,angle=0.0}
\caption{The observed rest-frame 
peak-energy as function of the inferred isotropic radiation energy for 
GRBs/XRFs of known redshift and well measured peak energy. The thick line 
is the best fitted power-law correlation, $E_p\sim 
[E_\gamma^{iso}]^{0.46\pm 0.04}\,.$ The dotted lines border the estimated 
spread (a factor of $\sim 4$) in the isotropic radiation energy due to the 
spread in $\gamma$ and the angular dependence of the Thomson cross 
section. The large outlying triangles, which represent $E_p$ in GRB 980425 
as inferred by Ghirlanda et al.~\cite{Ghir2004} and a lower limit on $E_p$ 
in GRB 031203 reported by Sazonov et al.~(\cite{Sazo2004}), may correspond 
to a second peak (see Dado \& Dar~\cite{Dada2005}).} 
\label{fig1} 
\end{center} 
\end{figure}

\noindent 
If core collapse SNe and their environments were all identical, 
and if their ejected CBs were also universal in number, mass, Lorentz 
factor and velocity of expansion, all differences between GRBs would 
depend only on the observer's position, determined by $z$ and the angle of 
observation, $\theta$. For a distribution of Lorentz factors that is 
narrowly peaked around $\gamma\simeq 10^3$, the $\theta$-dependence is in 
practice the dependence on $\delta$, the Doppler factor. Hence 
Eqs.~(\ref{Epeak1}),(\ref{Eiso}) yield the correlation (Dar \& De 
R\'ujula~\cite{DD2000}; \cite{DD2004}), 
\begin{equation} 
(1+z)\,E_p\propto [E^{iso}_\gamma]^k\, , 
\label{EisoEp} 
\end{equation} 
with $k=1/3$, in good agreement 
with $k=0.35\pm 0.06$ found by Amati~(\cite{Amat2004}) from an analysis of a 
sample of 22 GRBs, which were detected and measured with instruments on 
board BeppoSAX, CGRO and HETE-2, and whose redshift $z$ became available 
from ground based follow-up optical observations. However, $k=1/3$ 
is marginally consistent with $k=0.40\pm 0.05$ which was 
found by Ghirlanda et al.~(\cite{Ghir2004}) from a fit to all (40) GRBs 
and XRFs with known redshift before June 2004. But, GRBs are far from being 
standard candles and relation~(\ref{EisoEp}) with $k=1/3$ is only a crude 
approximation. For instance, for GRBs with a small viewing angle, 
$\theta^2\, \gamma^2\ll 1\,,$ Eqs.~(\ref{Epeak1}) and (\ref{Eiso}) imply 
$(1+z)\, E_p\propto \gamma^2$ and $E^{iso}_\gamma \propto \gamma^4\,.$ 
Then, the spread in $\gamma$ yields $k=0.5\, .$ In the CB model, the 
expected value of $k$ for the various samples of GRBs and XRFs varies 
between $0.33$ and $0.50$. Indeed, the best fitted power-law for $(1+z)\, 
E_p$ as function of $E^{iso}_\gamma$ for all GRBs/XRFs of known redshift, 
$E_p$ and $E_\gamma^{iso}\, ,$ shown in Fig.~\ref{fig1} by a thick line, 
has $k=0.46\pm 0.05$ (Dado \& Dar~\cite{Dada2005}). The parallel thin 
lines in Fig.~\ref{fig1} border the expected spread around the best fit 
because of the spread in the `standard candle' properties of GRBs which 
was found in the CB model (Dar \& De R\'ujula~\cite{DD2004}). As shown in 
Fig.~\ref{fig1}, the correlation predicted by the CB model of GRBs/XRFs is 
well satisfied, except by GRBs 980425 and 031203 where $E_p$ is much 
larger than expected from their $E_\gamma^{iso}\,.$ 
\\ 

\noindent{\bf 3.2 Spectrum and Spectral Evolution of GRB Pulses}\\ 
In the CB model, the predicted GRB spectrum from ICS of ambient 
light with a thermal bremsstrahlung spectrum by the electrons inside the 
CB, is, 
\begin{equation} 
{ dN_\gamma[1]\over dE} \propto 
\left({T_{eff}\over E}\right )^\alpha\; e^{-E/T_{eff}}+b\; 
(1-e^{-E/T_{eff}})\; {\left(T_{eff}\over E\right)}^\beta\, , 
\label{totdist} 
\end{equation}
where $ \alpha \approx 1\,,$ 
$\beta=(p+2)/2\approx 2.1\, ,$ $T_{eff}=\gamma\, \delta\, T/(1+z)$ and $b$ 
is a dimensionless constant. The values of $\alpha$ and $\beta$ may 
deviate from their indicated values, because the ambient radiation may 
deviate from a thin thermal bremsstrahlung, and the power-law index of the 
accelerated and knocked-on electrons after cooling may be larger than 
$p+1=3.3$ and increase with time. Also the ambient light temperature seen 
by the CB decreases with distance, approximately as 
\begin{equation} 
T_{eff}(t)\sim T_{eff}(0)\,\{1-exp[-(t_0/t)^2]\}, 
\label{Tevol} 
\end{equation} 
where $t_0$ is a constant. As was shown in Dar \& De 
R\'ujula~\cite{DD2004} and in Dado \& Dar~\cite{Dada2005}, 
Eq.~(\ref{totdist}) is practically indistinguishable from the 
phenomenological Band function (Band et al.~\cite{Band1993}) and it is in 
good agreement with the measured spectrum of the photons in the first peak 
of the spectral-energy-density of ordinary GRBs and XRFs.\\

\begin{figure}[h]
  \vspace{-70mm}
  \begin{minipage}[t]{0.5\linewidth}
  \centering
  \includegraphics[width=70mm,height=100mm]{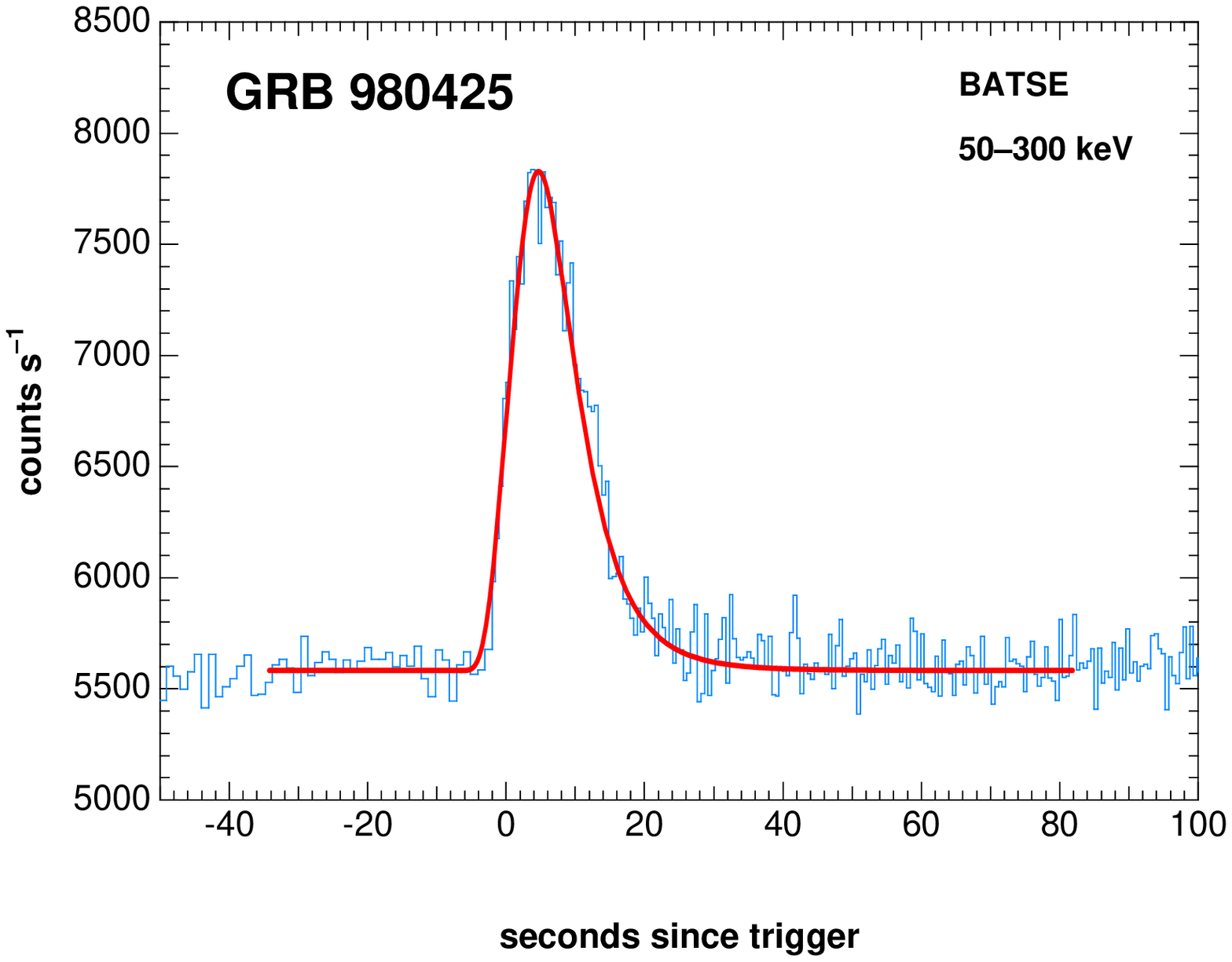}
  \caption{{\small The light curve $dN_\gamma/dt$ of GRB
   980425, as seen by BATSE in the 50--300 keV energy range
   and its CB model fit (Dar and De R\'ujula~\cite{DD2004}).}}
   \label{425p}
  \end{minipage}%
  \begin{minipage}[t]{0.5\textwidth}
  \centering
  \includegraphics[width=80mm,height=130mm]{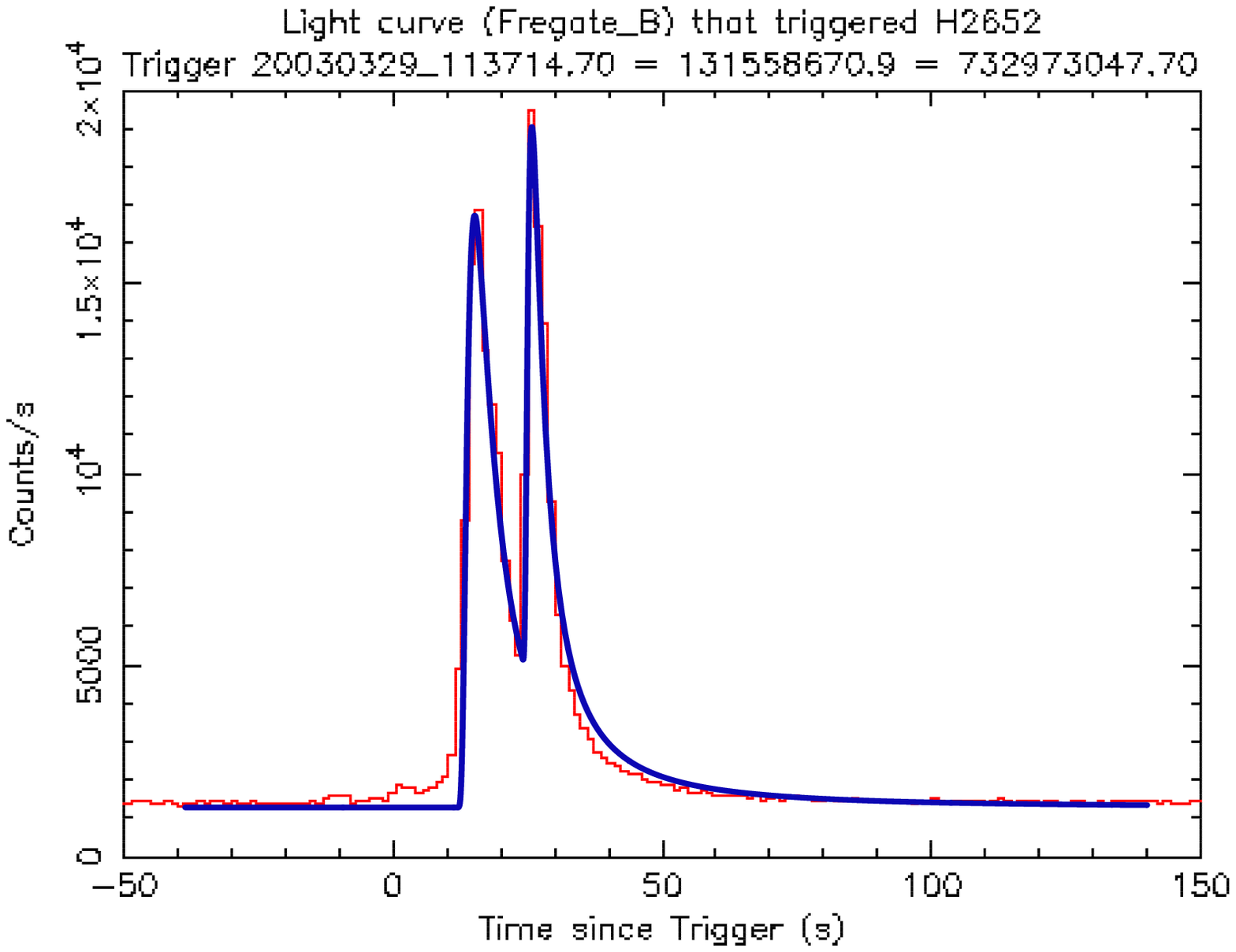} 
  \caption{{\small The lightcurve $dN_\gamma /dt$ of GRB 030329,
  as measured by HETE II showing two dominant
  pulses and the pulse shapes given by
  Eq.~(\ref{pheno}), where only the pulses' heights, widths and
  relative delay have been adjusted (see Dado et al \cite{DDD2003}).}}
  \label{329p} 
  \end{minipage}%
  \end{figure}

\begin{figure}
  \begin{center}
 \vspace{-40mm} \hspace{3mm} 
\psfig{figure=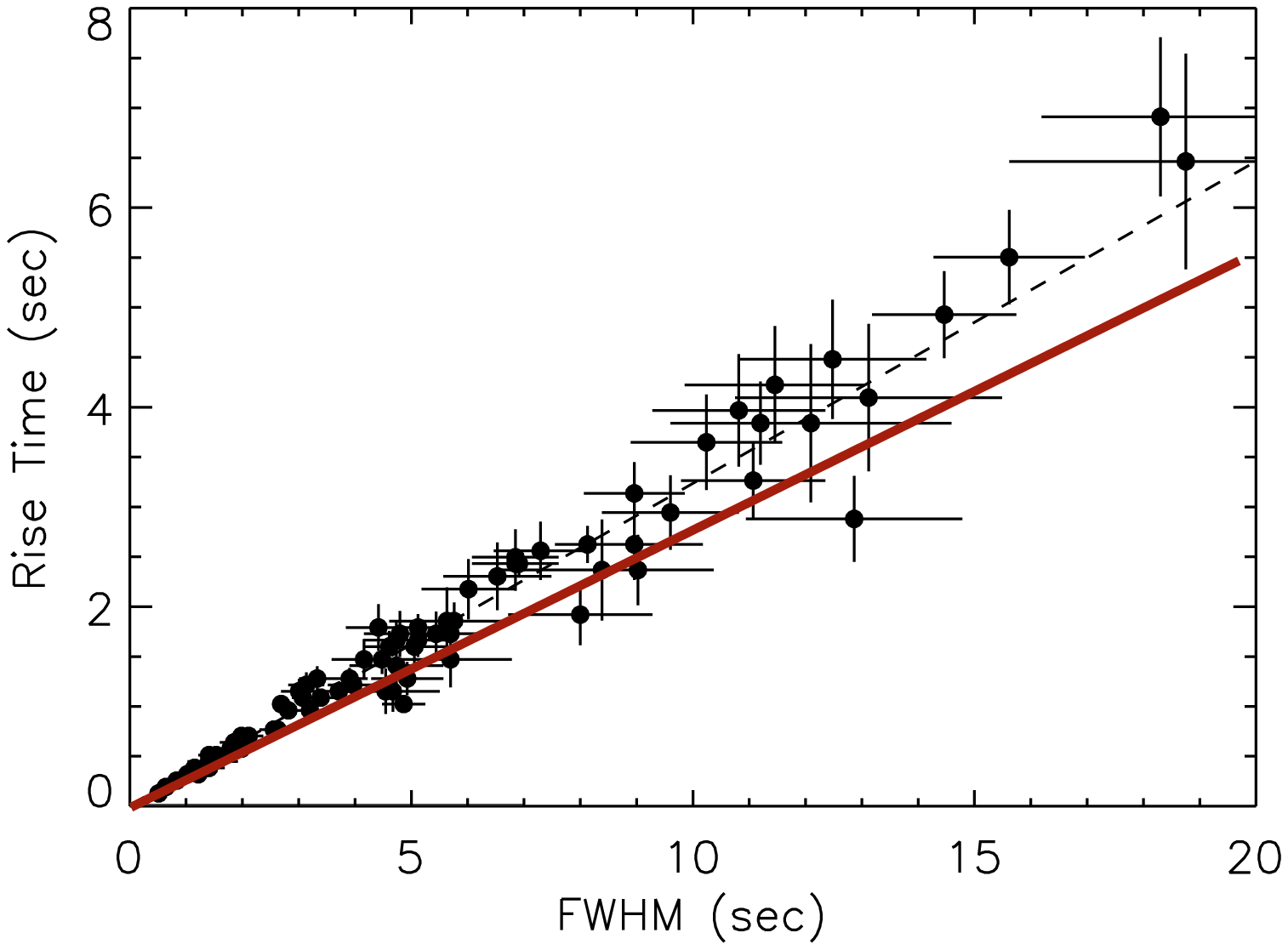,width=90mm,height=80mm,angle=0.0}
 \vspace{40mm} 
\caption{Rise-time (RT) from half-maximum to maximum versus 
full width at half-maximum of an ensemble of GRB single pulses (Kocevski 
et al.~2003). The data are from pulses of bright BATSE GRBs, the 
theoretical prediction (the continuous line) is from the naive pulse shape 
of Eq.~(\ref{pheno}). The dotted line is the best linear fit, RT=0.3 
FWHM. The figure has been adopted from Dar \& De R\'ujula~\cite{DD2004}).} 
\label{figFWRise} 
\end{center} 
\end{figure}

\noindent{\bf 3.3 Pulse Shape}\\ 
The CB model predicts a general shape of 
GRB pulses, 
\begin{equation} 
{dN\over dt}=exp\left[-\left({t_1\over t}\right)^m\right]\, 
       \left\{1-exp\left[-\left({t_2\over t}\right)^n\right]\right\}\, . 
\label{pheno} 
\end{equation} 
For instance, 
this pulse-shape fits very well the shapes of the famous
single-pulse GRB980425 
and the shapes of the two pulses of the famous GRB030329 as demonstrated 
in Figs.~\ref{425p} and \ref{329p}. 
Even the most naive pulse-shape, with 
$m=n=2$, $t_1=t_2$, shown in Figs.~\ref{425p} and \ref{329p}, does a very 
good job 
at describing individual ``FRED'' shapes as well as results averaged over 
all observed GRB shapes. For example, for this pulse-shape, the ratio of 
the rise-time from half-maximum to maximum to the total width at 
half-maximum is $\simeq 0.27$, while the observed result, reproduced in 
Fig.~\ref{figFWRise} is $\sim 0.3 $ (Kocevski et al.~\cite{Koce2003}). 

\noindent{\bf 3.4 Polarization}\\ 
The predicted polarization of ambient 
photons scattered by inverse Compton from the CB electrons into a viewing 
angle $\theta$ is 
\begin{equation} 
\Pi(\theta,\gamma)\approx {2\;\theta^2\,\gamma^2/(1+\theta^4\,\gamma^4)}, 
\label{polSN} 
\end{equation}
 which, for the probable viewing angles, $\theta\sim 
1/\gamma$, is naturally large (Shaviv \& Dar~\cite{SD1995}; Dar \& De 
R\'ujula~\cite{DD2004}).

\noindent 
Synchrotron radiation (SR) from a power-law distribution of 
electrons $dn_e/dE\sim E^{-p}$ in a {\it constant} magnetic field can 
produce a large polarization, $\Pi=(p+1)/(p+7/3)$, that is $\approx 70\%$ 
for a typical $p\approx 2.2$. But a collisionless shock acceleration of 
the electrons requires highly disordered and time varying magnetic fields 
(e.g. Zhang \& Meszaros~\cite{ZM2004}; for a dissenting view on this 
point, see Lyutikov, Pariev \& Blandford~\cite{LPB2003}). Consequently, 
the expected polarization of GRB and their afterglow is very small (see 
e.g., Medvedev and Loeb~\cite{ML1999}; Lyutikov, Parviev \& Blandford 
\cite{LPB2003}) if the 
$\gamma$-ray generating mechanism is synchrotron radiation from 
shock-accelerated electrons moving in highly entangled magnetic field. 
Indeed, this is the case for the measured polarization of GRB afterglows 
and the expected GRB polarization in the FB models. But a very large GRB 
polarization was reported for GRB 021206 (Coburn \& Boggs~\cite{CB2003}, 
see however Wigger et al.~\cite{Wigg2004}) and GRBs 930131 and 960924 
(Willis et al.~\cite{Will2005}), which clearly advocates the ICS of the CB 
model, as opposed to the SR of the FB models, as the mechanism generating 
the $\gamma$-rays of a GRB.

\noindent 
The reported large polarization of GRB 021206 prompted FB model 
papers on GRB polarization, which showed that under very contrived 
circumstances ---such as geometrical coincidences and unnaturally ordered 
magnetic fields--- shock models of GRBs may also produce a large linear 
polarization. In our opinion, this is what the articles on GRB 
polarization by Granot~(\cite{Gran2003});  Eichler \& 
Levinson~(\cite{EL2003}); Waxman~(\cite{Waxm2003b}); Nakar, Piran \& 
Waxman~(\cite{NPW2003}) show. Surprisingly, it is not what they say. For 
instance, Nakar, Piran \& Waxman~\cite{NPW2003} reach the opposite 
conclusion: {\it ``the recent detection of very high linear 
polarization... suggests strongly that these $\gamma$-rays are produced by 
synchrotron emission of relativistic particles.'' }

\section{FB and CB Models Confront Afterglow Data} 
\noindent In both, the 
FB models and the CB model, the late-time afterglow of GRBs is produced by 
synchrotron radiation from a power-law distribution of ISM electrons which 
were accelerated by the highly relativistic bipolar jets ejected in the SN 
explosion. However, the geometry and composition of the jet, the 
deceleration of the jet and the acceleration mechanism of the ISM 
electrons are different, and result in different predictions and 
interpretations of the AG data. Due to space limitation, I shall discuss 
only 4 key features of the AG phenomena:\\

\begin{figure}[h]
  \begin{minipage}[t]{0.5\linewidth}
  \centering 
  \includegraphics[width=75mm,height=80mm]{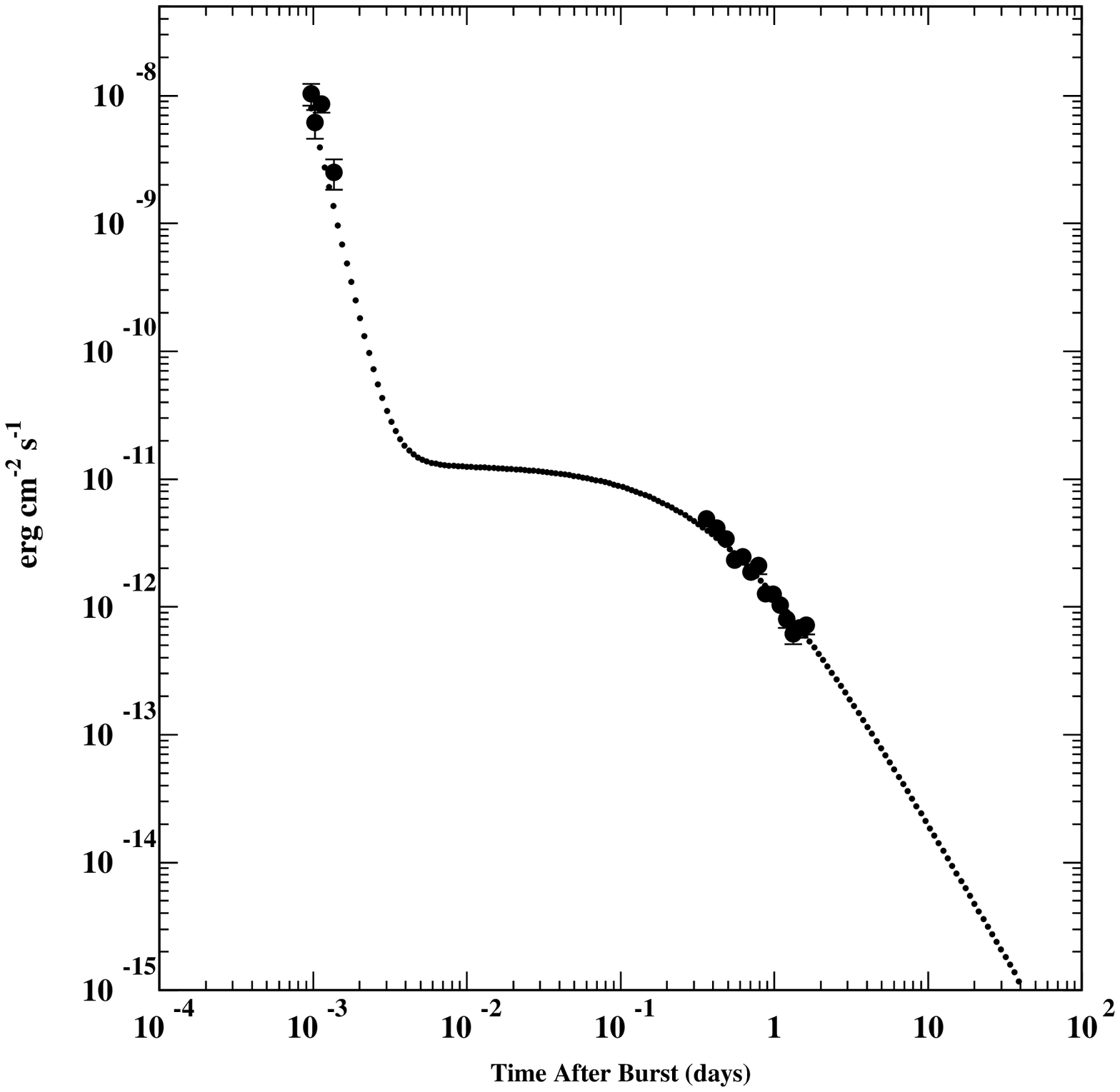}
  \caption{{\small The universal-shape of the X-ray afterglow of GRBs
  which was predicted by the CB model in 2002
  (Dado et al.~\cite{DDD2002}) compared with
  the early and late time X-ray
  afterglow (2-10 keV) of GRB 990510 as measured by
  BeppoSAX (Pian et al~\cite{Pian2001}).}}
  \label{510}
  \end{minipage}%
  \begin{minipage}[t]{0.5\textwidth}
  \centering
  \includegraphics[width=75mm,height=80mm]{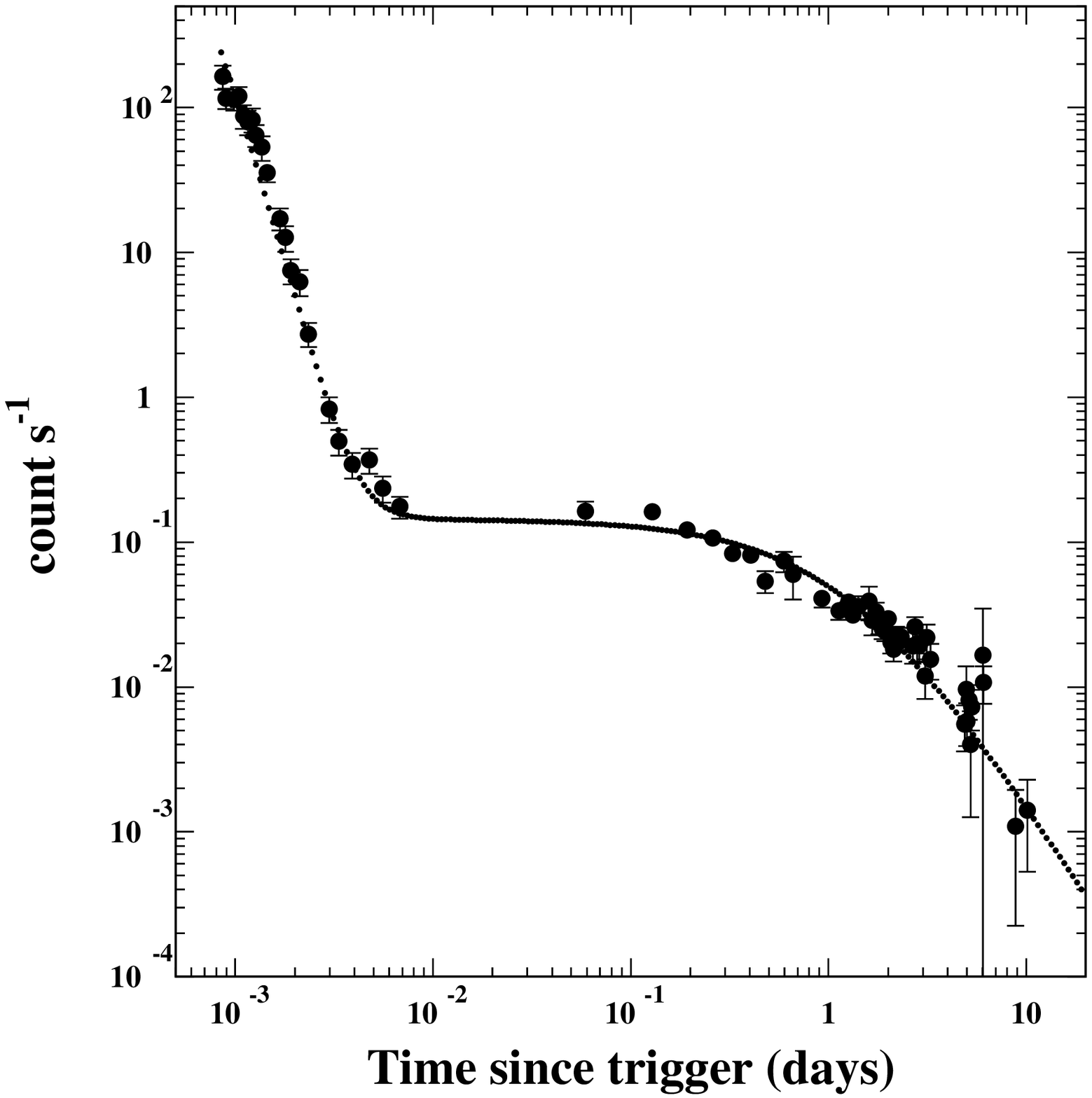} 
  \caption{{\small The X-ray afterglow (0.2-10 keV) of GRB 050315
 as measured by the 
  X-Ray Telescope (XRT) on board SWIFT (Vaughan et al.~\cite{Vaug2005})
  and a CB model fit (Dado et al. to be published).}} 
  \label{315} 
  \end{minipage}%
   \end{figure}

\noindent {\bf 4.1 The Canonical Shape of the Early X-ray Afterglows}\\ 
The early X-ray afterglows of GRBs measured by the XRT aboard SWIFT show a 
universal behaviour: the light curves broadly consist of three distinct 
power law segments: (i) an initial very steep decay ($t^{-\alpha}$ with 
$3<\alpha<5$ , followed by (ii) a very shallow decay ($0.2<\alpha<0.8$), 
which changes finally to (iii) a steeper decay ($1<\alpha<1.5$). These 
power law segments are separated by the corresponding ``ankle'' and 
``break'' times, $300s<t_{ankle}<500$s and $ 10^3s<t_{break}<10^5$s. This 
is demonstrated in Fig.~\ref{315} for GRB 050315 (Vaughan et 
al.~\cite{Vaug2005}). It was claimed that this universal behaviour cannot 
easily be explained by current theoretical models of GRBs. This may be 
true for the popular FB models of GRBs. It is not true, however, for the 
CB model which predicted this canonical behaviour (Dado et al.~ 
\cite{DDD2002}). This is demonstrated in 
Fig.~\ref{510} which is
borrowed from Dado et al.~(\cite{DDD2002}) 
where the predicted canonical 
behaviour was compared with observations of the X-ray AG of 
several GRBs
and in Fig.~\ref{315} with
the X-ray AG of GRB 050315 observed with the SWIFT XRT
(Vaughan et al.~\cite{Vaug2005}). 

\noindent
In the CB model, the energy flux per unit 
time due to bremsstrahlung and line cooling from a CB, after the CB 
becomes transparent to radiation ($t>t_{tr}$), seen by an observer at a luminosity 
distance $D_L(z)$, is given by, 
\begin{equation}
{dF\over dt}\approx
       {3\, \Lambda(T)\, N_b^2\, \delta^4
       \over 16\, \pi^2\, R^3\, D_L^2 }\, ,
\label{brem2}
\end{equation}
where $N_b$ is the CB's baryon number, $R$ its radius
and $\Lambda(T)$ its ``cooling function''.
If the loss-rate of the CB's internal energy is mainly adiabatic,
$T\!\propto\! 1/R^2$.  At a CB's transparency time, $\tau$,
$T\!\sim\! 10^4$--$10^5$ K  (Dado et al. \cite{DDD2002}), and $\Lambda(T)$ 
oscillates and
depends on composition in this $T$-range. A rough
description of the results of Sutherland \& Dopita \cite{Suth1993} is:
$\Lambda(T)\!\sim\! T^a$, with $a\!\sim\! 2$ for zero metallicity,
and $a\!\sim\! 0$ for high metallicity.
During the short TB+LE phase,
$\delta$ stays put and $R$ increases
approximately linearly with time.
The observer time $t$ is related to the CB's rest-frame
time $t'$ through $dt=(1+z)\, dt'/\delta$. Thus,
 $dF/dt\!\propto\! (t+\tau)^{-(3+2\, a)}$
and the expected powers are $\!\sim\! t^{-3}$ to $\!\sim\! t^{-7}.$
The TB+line
emission, which  decreases like $R^{-3}\, T^{\beta}\sim R^{-3+2\,a}$,
dominate the CBs' emission until synchrotron emission, which is 
proportional 
to $R^2,$ takes over. In the relevant ${\rm 10^4<T<10^5\, K}$ 
temperature range, 
on the average, $0.75<\beta<1.5$ for solar composition
composition, and TB+line 
emission from the CBs  
declines like $R^{-3}\, T^{a} \sim (t+tau)^{-(3+2a)}\sim t^{-5\pm 1} $  
(Dado et al.~\cite{DDD2002}). 

\noindent {\bf 4.2 Synchrotron AG from Decelerating Jets}\\ 
In the CB 
model, the jetted CBs, like those observed in $\mu$-quasars, are assumed 
to contain a tangled magnetic field in equipartition with the ISM protons 
which enter it. As the CBs plough through the ISM, they gather and scatter 
its constituent protons. The re-emitted protons exert an inwards pressure 
on the CBs which counters their expansion. In the approximation of 
isotropic re-emission in the CB's rest frame and constant ISM density, 
$n$, and a constant $\gamma$, one finds that within a few minutes of an 
observer's time, a CB reaches its asymptotic radius $R$. In the same 
approximation the deceleration of CBs in the ISM as function of observer 
time, which depends on the initial $\gamma=\gamma_0$ as they become 
transparent to radiation, on their baryon number $N_b$ and radius $R$, on 
the ISM density $n$ and on their viewing angle, $\theta\, , $ relative to 
their direction of motion, is given by, 
\begin{equation} 
[(\gamma_0/\gamma)^3-1]+3\,\theta^2\,\gamma_0^2\, [\gamma_0/\gamma - 1]
   = t/t_0\,,
 \label{cubic} 
\end{equation} 
where, $t_0 =(1+z)\, N_b/8\,c\, n\, 
\pi\, R^2\, \gamma_0^3\, .$ Hence, $\gamma$ and $\delta $ change very 
little with time as long as $t<t_{break}\, ,$ where, $t_{break}\approx 
(1+ 
3\,\theta^2\, \gamma_0^2)\, t_0\,,$ i.e., 
\begin{equation} 
t_{break}\sim 1.8\times 10^3\, (1+z)\,(1+ 3\, \theta^2\, \gamma_0^2)\, 
\left[{\gamma_0\over 10^3}\right]^{-3}\, 
\left[{n\over 10^{-2}\, cm^{-3}}\right]^{-1}\, 
\left[{R\over 10^{14}}\right]^{-2}\, 
\left[{N_b\over 10^{50}}\right]\, {\rm s} \, . 
\label{tbreakCB} 
\end{equation} 
In the CB model, the electrons that a CB gathers in its 
voyage through the ISM are Fermi-accelerated in the CB enclosed magnetic 
maze and cooled by synchrotron radiation to a broken power-law 
distribution with an {\it injection break} at the energy 
$E_b=m_e\,c^2\,\gamma(t)$ at which they enter the CB. Their emitted 
synchrotron radiation has a broken power-law form with a break frequency 
corresponding to $ E_b$. In the observer frame, before absorption 
corrections, it has the approximate form: 
\begin{equation} 
F_\nu \equiv \nu\, (dn_\gamma/ d\,\nu) \propto 
    n\, R^2\, [\gamma(t)]^{3\alpha-1}\, 
   [\delta(t)]^{3+\alpha}\, \nu^{-\alpha}\, , 
\label{sync} 
\end{equation} 
where $ \alpha\approx 0.5$ for $\nu\ll\nu_b$ and $ \alpha\approx p/2 
\approx 1.1$ for $\nu\gg\nu_b$, and 
\begin{equation}
 \nu_b \simeq 1.87\times 10^3\, [\gamma(t)]^3\,\delta(t)\, [n_p/ 
10^{-3}\;cm^{-3}]^{1/2}/(1+z)\, Hz. 
\label{nubend} 
\end{equation} 
The initial slow decline of $\gamma(t)$ and $\delta(t)$ produces 
the observed shallow decay of the early X-ray synchrotron
AG, provided that the density and the 
extinction of light from the CB are constant along the CB trajectory. The 
sum of TB emission and synchrotron emission from the CB produces the 
canonical X-ray light curve with an early fast TB decay, which is taken 
over at the ``ankle'' by synchrotron emission with an initial shallow 
decay that rolls over around the ``break'' to a steeper power-law decline, 
as demonstrated in Figs.~\ref{510} and \ref{315}. The shallow 
decline is not 
sensitive to the exact deceleration-law of CBs. For instance, if the CBs 
sweep in the ISM particles and continue to expand according to 
$R=R_0\,(\gamma/\gamma_0)^{-2/3}$ (Dado et al.~\cite{DDD2002}), then their 
deceleration-law becomes 
\begin{equation} 
[(\gamma_0/\gamma)^{8/3}-1]+4\,\theta^2\,\gamma_0^2\, 
[(\gamma_0/\gamma)^{2/3} - 1]= 8\,t/9\,t_0\,, 
\label{cubic2} 
\end{equation} 
and $t_{break}\sim (1+ 4\,\theta^2\, \gamma_0^2)\, (9/8)\,t_0$ is not 
significantly different from that given by Eq.~(\ref{tbreakCB}). 

\noindent
{\bf 4.3 The AG Break and the Frail relation}\\ 
The original blast-wave 
models assumed that GRBs and their afterglows are produced by spherical 
fireballs. The 1997 discovery of BeppoSAX that GRBs have afterglows that 
appear to decline with time like a single power-law was generally accepted 
as indisputable evidence in support of the model. However, spherical 
emission implies implausible energy release from small volumes. Repeated 
claims made by us in print since 1994 (e.g. Shaviv \& Dar~\cite{SD1995}), 
that cosmological GRBs and their afterglows 
(Dar~\cite{Dar1997},\cite{Dar1998}; Dar \& Plaga~\cite{DP1999}) are beamed 
emissions from highly relativistic jetted ejecta from stellar collapse, 
were olympically ignored.  GRB990123 with its record ``equivalent'' 
spherical energy release in $\gamma$ rays was the turning point of the 
spherical blast wave models. Fireballs became firecones 
(Rhoads~\cite{Rhoa1999}; Sari et al.~\cite{SPH1999}; Frail et 
al.~\cite{Frai2001}) or, more properly, firetrumpets, jets of material 
funneled in a cone with an initial opening angle (also called $\theta$) 
that increases as the ejecta encounter the ISM. For years these modellers, 
unaware of the Copernican revolution, placed us, the observers, at a 
privileged position, precisely on-axis, so that all detected GRBs would 
point exactly to us. More recently, the FB has evolved towards the 
realization that the observing angle {\it also} matters, a step in the 
right direction advocated by the CB model: the observation angle is the 
one that matters.

\noindent 
Conical FB are claimed to produce a break in the AGs when the 
beaming angle becomes larger than the opening angle of the conical jet, 
$\gamma(t)^{-1}\sim \theta_j\,, $ and an on-axis 
observer begins to 
see the full angular extent of the jet (e.g. Rhoads~\cite{Rhoa1999}; Sari 
et al.~\cite{SPH1999}). However, sharp breaks were not reproduced by 
detailed FB calculations that properly took into account arrival time and 
off-axis viewing effects and an assumed sidewise expansion of the conical 
jet on top of its ballistic motion, with a speed comparable to the speed 
of light (Sari et al.~\cite{SPH1999}). In order to avoid such a 
difficulties, observers usually use a free broken power-law 
parametrization (e.g., Stanek et al.~\cite{Stan1999}; Harrison 
et al.~\cite{Harr1999}) of the AG, which is fitted 
directly to the data rather than a properly derived AGs from the FB model.

\noindent 
In FB models which place the observer on/near the axis of the 
conical jet, the break time is given by (Sari et al.~\cite{SPH1999}), 
\begin{equation} 
t_{break}\sim 2.23\, (1+z)\left[{\theta_j\over 0.1}\right]^{8/3}
\left[{n\over 0.1\, cm^{-3}}\right]^{-1/3} 
\left[{\eta_\gamma\over 0.2}\right]^{-1/3} 
\left[{E_\gamma^{iso}\over 10^{53}\, ergs}\right]^{1/3}\, {\rm day} \, , 
\label{tbreakFB} 
\end{equation} 
where $\eta_\gamma$ is the efficiency of the FB in 
converting the energy of the ejecta into gamma rays and $n$ is the 
circumburst density.  Frail et al.~(\cite{Frai2001}) suggested that the 
$\gamma$-ray energy of conical GRBs is a standard candle independent of 
the opening angle of the conical GRB and consequently 
$E_\gamma^{iso}\approx E_\gamma\,\theta_j^2/4\,, $ which became known as 
the ``Frail Relation''\footnote{The fraction of the celestial sky which is 
lighted by a conical GRB is $f_b=(1-cos\theta_j)/2\approx \theta_j^2/4$ 
and not $(1-cos\theta_j)\approx \theta_j^2/2\,,$ which was used by Frail 
et al.~(2001) and in many following publications that made use of the 
``Frail Relation''.}. From analysis of 16 GRBs with known redshift, Bloom 
et al.~(\cite{Bloo2003}) found that $E_\gamma\approx 1.3\times 10^{51}$ 
ergs. The use of this value and the substitution 
$\theta_j^2=4\,E_\gamma/E_\gamma^{iso}$ in Eq.~(\ref{tbreakFB}), 
transforms it to, 
\begin{equation} 
t_{break}\sim 4.33\, (1+z) 
\left[{n\over 0.1\, cm^{-3}}\right]^{-1/3} 
\left[{\eta_\gamma\over 0.2}\right]^{-1/3} 
\left[{E_\gamma^{iso}\over 10^{53}\, ergs}\right]^{-1}\, {\rm day} \, . 
\label{tbreakFB2} 
\end{equation} 
However, all published attempts to use Eq.~(\ref{tbreakFB2}) to predict 
$t_{break}$ for AGs of GRBs from the observed $E_\gamma^{iso}$ before it 
was measured, have failed badly\footnote{For instance, Rhoads et 
al.~\cite{Rhoa2003} predicted $t_{break}>10.8$ days for GRB 030226 while 
shortly after, Greiner et al.~(\cite{Grei2003}) reported that 
 $t_{break}\sim 
0.8$ day.}, suggesting that the claimed success of the ``Frail 
relation'' may be an artifact and probably because of a posteriori 
adjustment of free parameters. Moreover, $E_\gamma^{iso}$ for all XRFs 
with known $z$ seems to be much smaller than the standard candle 
$E_\gamma\approx 1.3\times 10^{51}$ ergs, implying that XRFs and GRBs 
cannot be the same phenomenon viewed from different angles, contrary to 
other indications that they are (e.g. Dado et al.~\cite{DDD2004a}).\\

\begin{figure}[h]
  \begin{minipage}[t]{0.5\linewidth}
  \centering 
  \includegraphics[width=65mm,height=65mm]{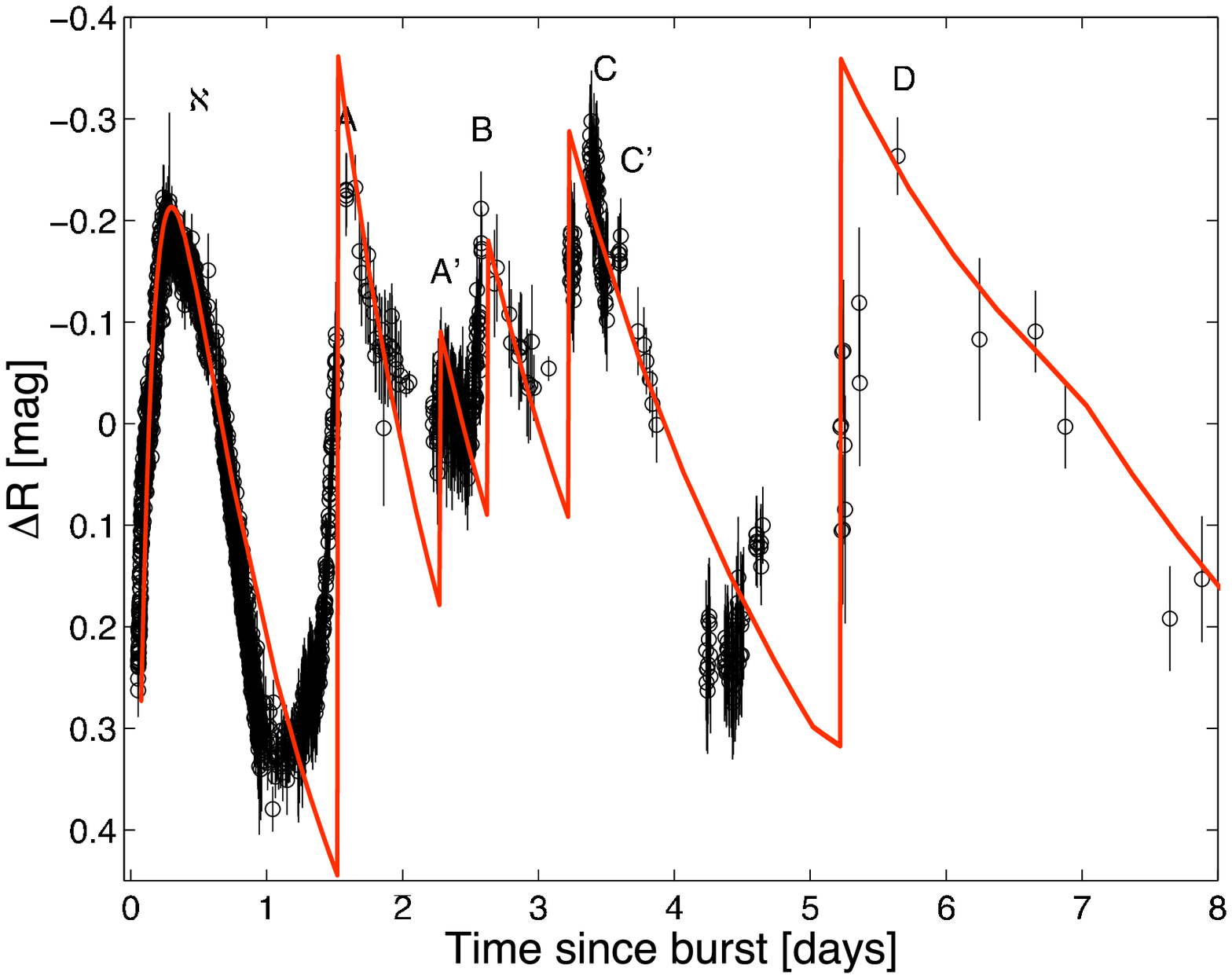}
  \caption{{\small Comparison between the R-band AG of GRB 030329, shown 
as ``residua" $\Delta R$ of the data (black points and circles) relative 
to a broken power law of index $-\alpha$ jumping from $\sim 1.1$ to $\sim 
2$ at $t\sim 5$ days (Lipkin et al.~\cite{Lipk2004}) and the residua, 
relative to the same broken power law, calculated from the CB-model (red 
line)  for the input density profile shown on the right hand side (Dado et 
al.~\cite{DDD2004b}).}}
\label{329}  
\end{minipage}%
  \begin{minipage}[t]{0.5\textwidth}
  \centering
  \includegraphics[width=65mm,height=65mm]{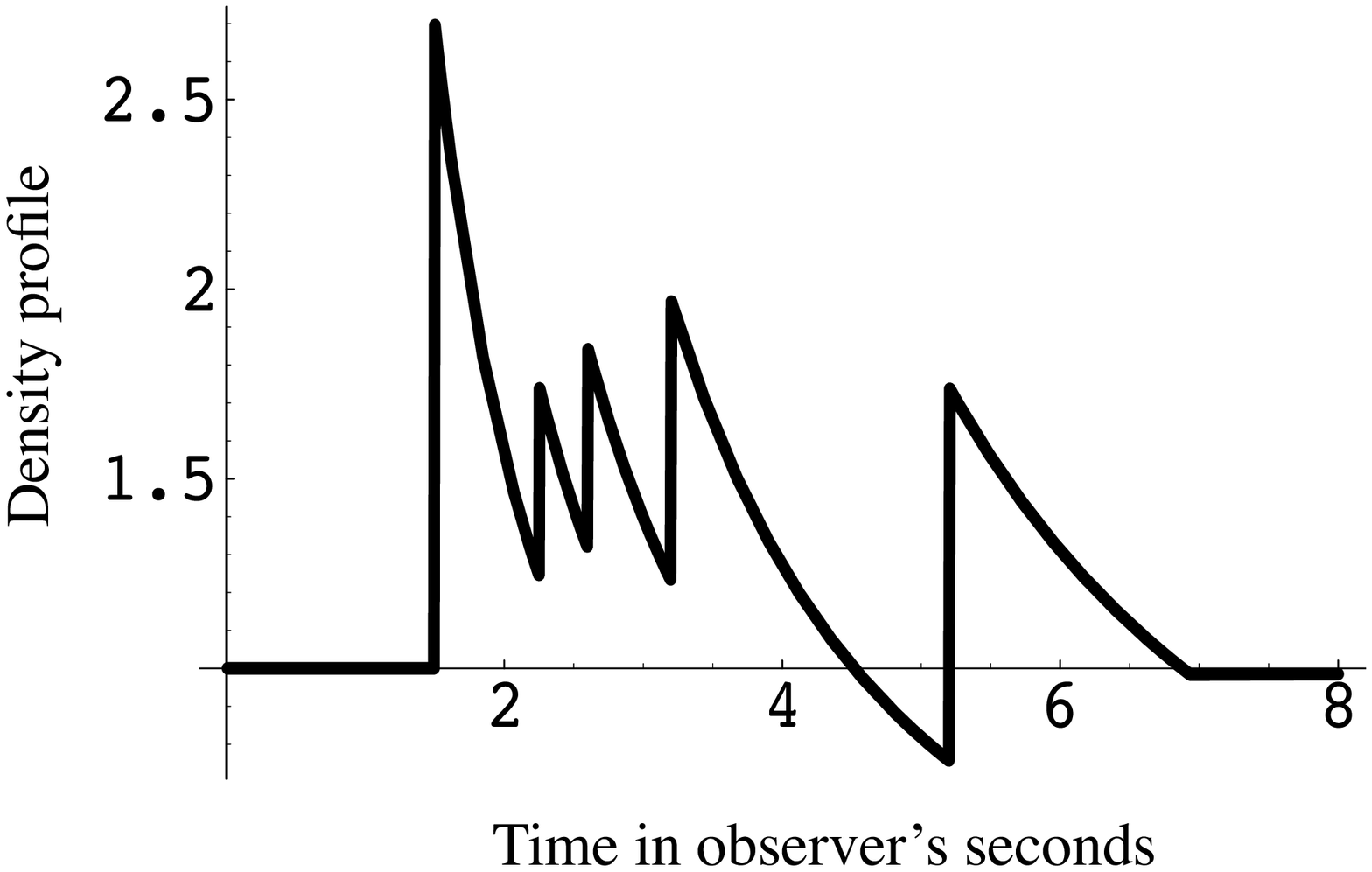} 
  \caption{{\small The density profile assumed in the CB-model fit to the 
R-band AG of GRB 030329 (relative to a smooth ISM density -- a constant 
plus a ``wind" contribution decreasing as $ 1/r^2$).}}
  \label{329d} 
  \end{minipage}%
\end{figure}

\noindent{\bf 4.4 AG Bumps and Flares}\\ 
The SWIFT XRT provided an unprecedented look at 
the behaviour of X-ray AG of GRBs in the first few hours after burst. 
While most of the early AGs have smoothly declining lightcurves, a 
substantial fraction has large X-ray flares on short time scales (see, e.g 
Burrows et al.~\cite{Burr2005}). In the CB model, 
such flares may result from late time accretion episodes on the compact
central object.  Late bumps and flares in the AG of GRBs 
have 
been detected before in the the X-ray AG of GRB 970508 and also in the optical 
AG of e.g. GRBs 000301c and 030329 and in the radio AG of GRB 030329. In 
the CB model, the AG is a direct and {\it quasi-local} tracer of the 
density of the ISM through which a CB travels: spatial changes in density 
translate into temporal changes in fluence.
In particular, large flares are expected due 
to CB encounters with density jumps along the CB trajectory (e.g. Dado et 
al.~\cite{DDD2002},\cite{DDD2004b}). Such density jumps are produced
in the circumburst environment by the
SN and pre SN ejecta, and by the winds in the 
ISM inside the superbubbles (SBs) where most core collapse SN take 
place, and in particular, at the complex SB boundaries created by stellar 
winds and previous supernovae in the SB. This is demonstrated in 
Figs.~\ref{329} and \ref{329d}. 
\\ 

\noindent
In the FB model, both early and late time flares result from 
extended central activity (e.g. Granot et al.~\cite{Gran2003}). 
Although such an activity can neither be predicted nor ruled out, it is 
not clear why such late time activity does not produce gamma ray pulses 
and why the duration and magnitude of the AG flares scale roughly with 
the time and the magnitude of the declining AG.

\section{Conclusions} 
\noindent 
Confronting the predictions of the FB and 
CB models with observations of GRBs and their afterglows unmistakenbly 
demonstrates that the CB model is incredibly more successful than the 
standard FB or blast-wave models of GRBs, which suffer from profound 
inadequacies and limited predictive power. The CB model is falsifiable in 
its hypothesis and results. Its predictions are summarized in simple 
analytical expressions, derived, in fair approximations, from first 
principles. It provides a good description on a universal basis of the 
properties of long-duration GRBs and of their afterglows (AGs). Therefore 
it is not surprising that the FB models have gradually progressed in the 
direction of the CB model (compare for instance, 
Piran~\cite{Pira1999},\cite{Pira2000},\cite{Pira2005}): 

\begin{itemize} 
\item 
The discovery (Stanek et al.~\cite{Stan2003}; Hjorth 
et al~\cite{Hjor2003}) of SN2003dh ---associated with GRB 030329--- on the 
8$^{\rm th}$ of April, 2003 as predicted a few days earlier by the CB 
model (Dado et 
al.~\cite{DDD2003}), has transformed the GRB-SN association, which 
underlines the CB model, from a minor and doubtful issue (e.g.~Hurley et 
al.~\cite{HSD2002}; Waxman~\cite{Waxm2003a}) into something crucial that 
{\it everybody always knew}.

\item 
The angular spread of the ejecta has progressively diminished from an 
original $4\,\pi$ solid angle, to a jet opening angle of tens of degrees 
(Frail et al.~\cite{Frai2001}), to $\theta_j=10$ mrad 
(Waxman~\cite{Waxm2003b}).

\item 
Following the CB model, the observer's angle, once upon a time set 
to zero by fiat in the FB model (e.g.~Rhoads~\cite{Rhoa1997, Rhoa1999}; 
Sari et al.~\cite{SPH1999}; Frail et al.~\cite{Frai2001}) is gaining a 
non-negligible role in FB models (see e.g. Granot et al.~\cite{GRP2005} 
and references therein).

\item 
The correlations discussed in Dar \& De R\'ujula 
(\cite{DD2000,DD2004}) should be approximately valid for any jets seen 
off-axis. No doubt these correlations will soon be fully exploited as 
a success of forthcoming off-axis FB models.

\item{} 
With their current, rather small $\theta_j$ values in the conical 
FB models, GRBs, once, in the era of the spherical FB models, 
systematically publicized  as {\it ``the biggest 
explosions after the Big Bang''}, have become a small fraction of the SN 
explosion energy, that is, what they always were in the CB model.

\end{itemize}

\noindent
It is not at all inconceivable that the FB models may continue to 
incorporate and ``standardize'' other aspects of the CB model. Four large 
stumbling blocks lie along this path:

\begin{itemize}
\item
The traditional GRB-generating mechanism in FB models, 
synchrotron emission from shock accelerated electrons in entangled 
magnetic fields, must be replaced by inverse Compton scattering 
of ambient light, as suggested, for instance, by 
the very large linear polarization measured in three bright GRBs. 

\item
The scenario of internal and external shocks powering, 
respectively, the GRB and AG must be forsaken:
The observed GRB energy is much larger than the AG energy. 
Energy-momentum conservation implies that the center of mass energy in 
shell collisions is much smaller than in their collision with the ISM, 
unless one assumes that most of the bulk motion kinetic energy 
of the jet is radiated in the internal collisions, which is not
supported by observations of relativistic jets in other systems. 

\item
The substance of which the FB-models' ejecta are made: a 
delicately baryon-loaded (that is, highly fined-tuned) plasma of $e^+e^-$ 
pairs. Such a fancy substance may not be so difficult to forsake, in 
comparison with good old ordinary matter. 

\item 

The main obstacle may be the magic wand of FB models: 
shocks. 
\end{itemize}
\noindent
If and when these obstacles are overcome, the fireballs may turn 
out to have always been cannonballs for, after all, in the CB model,... 
SNe fire balls.

\noindent
One serious drawback of the CB model is that it makes GRBs become very 
uninteresting, in comparison with what they used to be: one of the biggest 
mysteries of astrophysics and the biggest of explosions after the Big 
Bang. Fortunately, and independently of the ``peripheral'' GRB- and 
AG-generating physics, the biggest conundrum remains: How does an SN 
manage 
to sprout mighty jets? In the CB model the guidance along this path is 
better than simulations: the CBs responsible for GRBs are akin to the 
increasingly well--studied ejecta of quasars and microquasars presumably 
fired in hyper-accretion episodes on compact central objects. Moreover, the 
CB model underlines a unified theory of high energy astrophysical 
phenomena (e.g De R\'ujula~\cite{ADR2004a}) and bridges GRBs to other 
observational fields as well: cosmic rays (Dar \& Plaga~\cite{DP1999}; 
Dar~\cite{Dar2004},\cite{Dar2005}; De R\'ujula~\cite{ADR2004b}), the gamma 
background radiation (Dar \& De R\'ujula~\cite{DD2001a}), intergalactic 
magnetic fields (Dar \& De R\'ujula~\cite{DD2005}) and astrobiology (Dar 
et al.~\cite{DLS1998}; Dar \& De R\'ujula~\cite{DD2001b}).

\begin{acknowledgements}
This research was supported in part
by the Asher Fund for Space Research at the Technion.
It is based on an ongoing collaboration with S. Dado and A. De R\'ujula
for which the author is very grateful. S. Vaughan has kindly provided us
a tabulated measurememts of the X-ray AG of GRB 050315 with SWIFT XRT. 
\end{acknowledgements}

\end{document}